\documentclass[notitlepage]{revtex4-1}

\usepackage{amsthm}
\usepackage{amsmath}
\usepackage{graphicx}
\usepackage[table]{xcolor}
\usepackage{colortbl}
\usepackage{color}

\theoremstyle{plain}

\theoremstyle{definition}

\theoremstyle{remark}

\newcommand{\del}[1]{}

\begin{document}

\title{Afterpulsing studies of low noise InGaAs/InP single-photon negative feedback avalanche diodes}

\author{Boris Korzh}
\thanks{Corresponding author email: boris.korzh@unige.ch}
\author{Tommaso Lunghi}
\author{Kateryna Kuzmenko}
\author{Gianluca Boso}
\author{Hugo Zbinden}
\affiliation{Group of Applied Physics, University of Geneva, Geneva, Switzerland}
\date{October 2014} 

\begin{abstract}

We characterize the temporal evolution of the afterpulse probability in a free-running negative feedback avalanche diode (NFAD) over an extended range, from $\sim$300~ns to $\sim$1~ms. This is possible thanks to an extremely low dark count rate on the order of 1~cps at 10\% efficiency, achieved by operating the NFAD at a temperatures as low as 143~K. Experimental results in a large range of operating temperatures (223-143~K) are compared with a legacy afterpulsing model based on multiple trap families at discrete energy levels, which is found to be lacking in physical completeness. Subsequently, we expand on a recent proposal which considers a continuous spectrum of traps by introducing well defined edges to the spectrum, which are experimentally observed. 

\end{abstract}
\maketitle

\section{Introduction}

Single photon detectors (SPDs) at telecom wavelengths are important for many applications including general quantum optics experiments, quantum key distribution~\cite{Gisin2002} (QKD), optical time domain reflectometry~\cite{Eraerds2010}, photon-counting optical communication~\cite{Robinson2006}, eye-safe laser ranging~\cite{Warburton2007} and biomedical imaging~\cite{Gemmell2013}. 

It is often convenient to operate SPDs in the free-running regime, where gating is not required. Superconducting nanowire SPDs (SNSPDs) generally provide the ultimate performance in this regime due to the possibility of achieving very high efficiency~\cite{Marsili2013}, low noise~\cite{Shibata2013}, high count rates and low temporal jitter~\cite{You2013}. However, due to the cryogenic temperature requirement~\cite{Verma2014} these detectors are deemed inconvenient and expensive for some applications.  InGaAs/InP single-photon avalanche diodes (SPADs)~\cite{Itzler2011} are a cheaper and less demanding alternative for these wavelengths, hence they play a crucial role in many technologies. 

Unfortunately, developing free-running detectors based on InGaAs/InP is challenging due to afterpulsing effects, where spontaneous dark detections can occur shortly after a photon detection. This arises due to charge carrier trapping at defect sites in the SPAD multiplication region, the subsequent release of which can lead to another avalanche. One of the main techniques used for minimizing the afterpulsing effect is the reduction of the avalanche current, since this reduces the probability that a defect site gets filled in the first place. This is generally achieved by operating the detectors in the gated regime, using short gates of less that a nanosecond~\cite{Yuan2010, Walenta2012, Restelli2013}. However, in order to operate in the free-running mode, the avalanche has to be either actively or passively quenched~\cite{Cova1996}. Passive quenching is typically achieved through the use of a large resistor placed in series with the SPAD. During a detection, as the avalanche current builds in the SPAD, it flows through the resistor, lowering the voltage across the SPAD until it is below the breakdown voltage, quenching the avalanche. To reduce any parasitic capacitance introduced by the series resistor, it needs to be monolithically integrated directly on the surface of the SPAD. Such devices are know as negative-feedback avalanche diodes (NFADs)~\cite{Itzler2009a}. 

The benefit of passive quenched NFADs is that they can be operated under a DC bias voltage, making free-running operation possible~\cite{Lunghi2012, Yan2012}. We have recently demonstrated that these detectors can achieve extremely low dark count rates (DCR), as low as 1~cps at 10\% efficiency~\cite{Korzh2014a}. Such values of DCR approach those only typically possible with SNSPDs~\cite{Eisaman2011}, which opens up the door to new applications previously not thought feasible with InGaAs/InP SPADs. For example, we have shown that these detectors can be used for long distance QKD~\cite{Korzh2014a, Korzh2014c}, increasing the maximum transmission distance by a factor of two, compared to previous InGaAs/InP SPAD based demonstrations. 

The low DCR of the NFAD detectors was achieved by lowering the operating temperature down to around 160~K, using a Stirling cooler, and introducing an active hold-off time to reduce the afterpulse probability (AP) further~\cite{Lunghi2012, Korzh2014a}. An extensive study of the NFAD behaviour at low temperatures was subsequently carried out to understand the effects on the DCR, afterpulsing and temporal jitter~\cite{Korzh2014a, Korzh2014b}. 

In this article, we begin by reviewing the main characteristics of NFAD operation at low temperatures. We then move on to concentrate on afterpulsing in more detail. More specifically we aim to study the temporal dependence of the AP. Such studies are of importance in understanding the microscopic origin of the afterpulsing, which could aid the mitigation of these effects in the future. Indeed, very little is know about the trap sites in the predominantly used InP material~\cite{Itzler2012} moreover a complete model which reproduces the observed afterpulse behavior is also missing. Thanks to the ability of achieving extremely low dark count probabilities by cooling the NFADs, we have the opportunity to study the evolution of the AP over a previously unattainable temporal range from $\sim$300~ns to almost 1~ms. To gain physical understanding of the experimental results we describe the legacy afterpulsing model based on a few trap-families at discrete activation energies. We show that such a model does not agree well with the experimental results and we go on to provide considerations for an extended mathematical model to describe the observed behavior.

\section{Low temperature operation of NFADs}

In this section we provide a short review of the main characteristics of the NFAD operation at low temperatures, down to $\sim$160~K. This includes the DCR, temporal jitter and afterpulsing. 

\subsection{Dark count rate and quantum efficiency}

The DCR is one of the most important metrics for a single photon detector. In Ref.~\cite{Korzh2014a} we showed for the first time that it is possible to operate NFADs with a DCR as low as 1~cps at 10\% efficiency, by operating them at 163~K. In Ref.~\cite{Korzh2014b} we presented the DCR dependence on temperature between 160~K and 220~K. In this range, it was apparent that two different dominating mechanisms were responsible for the dark-carrier generation. 

NFADs have a separate absorption and multiplication (SAM) structure, where InGaAs is used for the absorption region and InP with its larger bandgap is used for the high-field multiplication region. Dark-carrier generation can occur due to either a field dependent tunneling process, specifically trap-assisted tunneling (TAT), in the multiplication region (InP) or a thermally driven process in the absorption region (InGaAs)~\cite{Itzler2011}. We found that above 200~K, thermally generated carriers in the absorption region were dominant, whilst below 200~K, TAT in the multiplication region overtook. TAT itself is expected to have a small temperature dependence, however the DCR still exhibited an exponential decrease with falling temperature below 200~K. This arises due to the reduction in the breakdown voltage, leading to a reduction of the electric field required for a given detection efficiency.

At 163 K, a DCR of 1.2 cps was obtained~\cite{Korzh2014a} for a detection efficiency of 11.6\% (1550~nm), which is nearly two orders of magnitude better compared to the previously demonstrated record for free-running InGaAs/InP detectors~\cite{Yan2012}. At 27\% efficiency the DCR was about 15 cps.

Another consideration is the fact that the absorption coefficients in InGaAs are also temperature dependent. Indeed, the upper cut-off wavelength shifts by about 50~nm per 75~K~\cite{Acerbi2013}. This means that, for 1550~nm, the absorption efficiency in InGaAs is expected to drop-off sharply below 150~K. Therefore it is not expected that decreasing the temperature further will yield any benefit in terms of signal to noise ratio, unless of course a shorter wavelength is of interest.  

\subsection{Temporal Jitter}
  
Timing jitter is a very important figure of merit for a single photon detector, especially for high-speed applications, where high temporal resolution is crucial. It is defined as the temporal uncertainty of the detection announcement with fixed arrival time of photons. The dominant cause of the timing jitter is the stochastic nature of the impact ionization process that produces the detection avalanche~\cite{Itzler2011}. Once the avalanche is triggered, a fundamental build-up time is required for the avalanche amplitude to reach a predetermined threshold level. Crossing this threshold signals the detection event via the readout electronics. The build-up time and its standard deviation, here referred to as the timing jitter, both decrease with increasing detection efficiency~\cite{Tan2007}. In Ref.~\cite{Korzh2014b} we found that the jitter reduced at lower temperatures, for all detection efficiencies. For example, at 22\% efficiency the jitter improved from 185 ps to 155 ps by reducing the temperature from 223 K to 183 K. It was predicted that this dependence originates from the increase of the electron and hole ionization coefficients in the InP multiplication region with the reduction of temperature~\cite{Zappa1996}, which reduces the avalanche build-up time.

\subsection{Afterpulsing}

The total AP in SPADs increases exponentially with lower temperature. This is due to the increase in the trap lifetime, meaning that at lower temperatures trap families with shallower energy levels start to contribute to the total AP. For example, with a dead-time of 20~$\mu$s and a detection efficiency of 11.5\%, the total AP increases from about 0.1\% to about 2.2\% if the operating temperature changes from 223~K to 163~K~\cite{Korzh2014b}. Since this AP is acceptable for many applications, one can see that it is feasible to operate NFADs at low temperatures with extremely low DCR. 

One particularly interesting finding in Ref.~\cite{Korzh2014b} was that the temporal evolution of the AP followed a power-law dependency, which has been predicted by a model recently proposed in Ref.~\cite{Itzler2012}. The suggested physical significance of this is that there could be a broad spectrum of trap levels with a specific distribution of de-trapping rates. This is contrary to the legacy belief, that one, or just a few families of defects with exponential carrier de-trapping rates dictate the afterpulsing behavior. Although the power-law dependence seemed to agree with the proposal in Ref.~~\cite{Itzler2012}, the slope of the decay curves was found to be different. The reasons for this were not conclusive, since the investigation was carried out in different conditions, i.e. the temporal and temperature ranges were different. It is these discrepancies that has motivated the investigations outlined in this article.

\section{Afterpulsing model with discrete trap energy distribution}\label{sec:model1}

The most widely accepted model found in literature to describe the temporal distribution of the AP accounts for the existence of a limited number of trap families in the semiconductor, each with a distinct release lifetime. This model can be described as follows:
\begin{equation}\label{eq1}
p_{ap}(t) = \sum_{i=1}^N A_i\frac{1}{\tau_i(E_i, T)}e^{-\frac{t}{\tau_i(E_i, T)}}
\end{equation}
where the index $i$ labels the family of traps and $A_i$ is the probability of a trap being filled during an avalanche event. Here we considered the general case where  $N$ is the total number of trap families included in the model (typically $N\leq5$).

The lifetime of a trap, $\tau_i(E_i, T)$, is defined by the following formula:
 \begin{equation}\label{eq2}
 \tau_i(E_i,T) = \tau_0e^{\frac{E_i}{kT}} 
 \end{equation}
where $E_i$ is the activation energy of the trap (i.e. the difference between the conduction band and the trap level), $\tau_0$ is a parameter related to the carrier type and the band structure, $k$ is the Boltzmann constant and $T$ is the temperature.

\begin{figure}
\begin{center}
\includegraphics[width=0.6\textwidth]{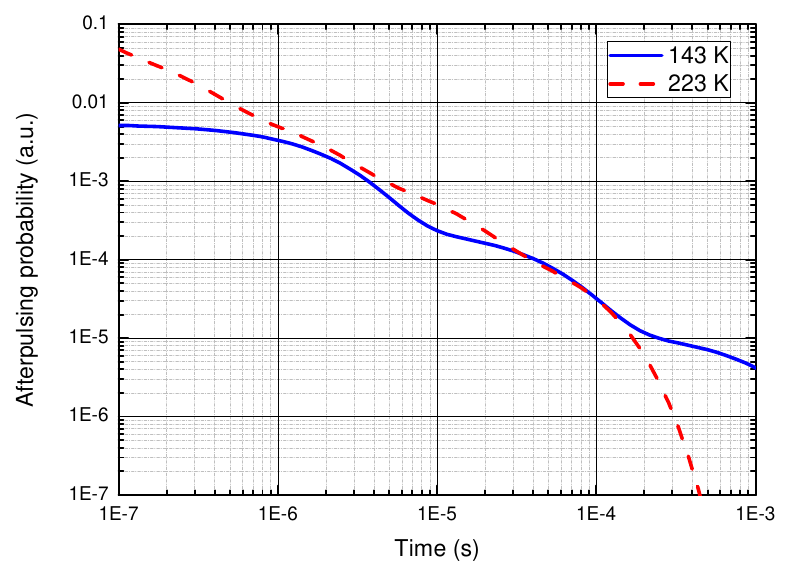}
\end{center}
\caption{Numerical simulation of the temporal distribution of the AP in a device with 5 discrete trap families with activation energies, $E_i$ = 0.15, 0.1875, 0.225, 0.3~eV. The other parameters were fixed to $\tau_0 = 10^{-11}$~s and $A_i = 10^{-8}$.}\label{fig:simulation}
 \end{figure}

By exploiting the temperature dependence of the lifetime of each trap family, we can test the discrete trap model by acquiring the temporal distribution of the AP at different temperatures. Figure~\ref{fig:simulation} shows a numerical simulation of Eq.~\ref{eq1}. In this simulation we have considered 5 trap families equally distributed between 0.15 and 0.3 eV.
Note that the bandgap of InP is $\sim$1.35~eV, hence these traps are located near the conduction band. At the higher temperature (223~K), the lifetimes of the traps, given by Eq.~\ref{eq2}, are close enough such that the individual decay curves cannot be resolved. The resulting curve can be described with a power-law trend up to approximately 100~$\mu$s as proposed in Ref.~\cite{Itzler2012}. On the other hand, by lowering the temperature down to 143~K, the difference between two adjacent trap lifetimes increases. When the lifetimes are well spaced the afterpulsing curve shows a step-like trend in time. Note that, if the trap levels are discrete, there will always be a temperature at which the individual lifetimes can be distinguished. In fact, one can compute the ratio between the characteristic lifetimes of two different energy levels as:
 \begin{equation}
\tau_{i+1}/\tau_{i} = e^{\frac{\Delta E}{kT}}
\end{equation}\\
For the proposed temperature of $T=143$~K and a spacing of one order of magnitude between the lifetimes, the maximum spacing between energy levels, which would not give rise to noticeable step-like behavior would be only $\Delta E = 0.03$~eV.

In addition, it is interesting to note that by exploiting the temperature dependence of the de-trapping rate, it is possible to probe the characteristic lifetime of the deepest trap (slowest de-trapping rate) and the shallowest trap (fastest de-trapping rate). In the case of our simulation, the lifetime of the shallowest trap is $\sim 2~\mu$s at the lowest temperature, causing the afterpulsing curve to be purely exponential in a range smaller than 1~$\mu$s. For the same reason, at the high temperature, after 100~$\mu$s, only the deepest trap is populated i.e. after this time the AP depends solely on its lifetime.

Experimental evidence presented in our previous work~\cite{Korzh2014a} showed that, even at low temperatures (down to 163 K), the AP can still be fitted with a power-law trend and that no step-like behavior is evident, suggesting that the discrete trap model gives only a partial description of the afterpulsing effect. Moreover, the time-resolved acquisition of the AP over an extended temperature (down to 143 K) and time range (up to 1 ms) has never been reported in literature. In the proceeding sections, we will present our experimental results from which a different physical model can be considered.

\section{Experimental setup}\label{sec:exp_setup}

Characterization of the aftepulsing in SPADs involves the measurement of the correlation of dark count events occurring after a photon detection. When SPADs are operated in the gated mode, characterization of the AP is usually done by the well-known double-window method~\cite{Cova1991}.  In this case, straightforward variation of the hold-off time between two gates achieves the characterization. However, in the free-running regime the procedure is more complex. 

In Ref.~\cite{Lunghi2012} we introduced a characterization method for the AP employing a field programmable gate array (FPGA). The FPGA first imposes a cycle where it waits for no detection to occur in a user defined time, which in our case can be chosen between 20~ns and 200~$\mu$s, to ensure that the detector is in a well defined initial condition. Once this condition is fulfilled, a pulsed laser is triggered and the probability of a photon detection in the corresponding time-bin provides a direct measure of the efficiency. Please see Refs.~\cite{Lunghi2012, Korzh2014a} for a detailed description of the detection efficiency characterization. Conditioned on a detection of the laser pulse, the FPGA looks for an afterpulse and subsequently updates a temporal histogram. Unlike the normal double-window method, this procedure allows the higher-order afterpulse contributions to be easily characterized. During the characterization, the user can choose to plot only the first afterpulse detection or to include all of the detections. 

During the afterpulse data acquisition, the FPGA sorts the detection events into 250 time-bins and the user can change the temporal range by choosing how many FPGA cycles are grouped into a single time bin. With a clock frequency of 50~MHz, we have a maximum temporal resolution of 20~ns, whilst hold-off times on the order of ms could also be investigated by grouping many FPGA cycles together. With this setup, in a single measurement run, we can collect the AP decay for just over one order of magnitude difference in time. In order to increase the total range of the AP characterization we carry out several measurement runs with different dead-time settings. 

In this work, we are concerned with studying the temporal evolution of the AP, therefore we chose to record only the first afterpulse following a photon detection. Because of this, we need to correct for pile-up distortion effects, which become apparent at long temporal delays. More precisely, the probability density function (pdf), of a noise detection event following a photon detection is defined as,
\begin{equation}
p(i)= \frac{C_i}{C_\text{d}-\sum\limits_{j=0}^{i-1}{C_j}},
\label{eq:pdf}
\end{equation}
where $C_i$ is the count rate in time bin $i$ and $C_\text{d}$ is the total number of laser detections. The $C_j$ term accomplishes pile-up correction. 

In order to isolate the AP from this pdf, one needs to characterize the intrinsic DCR. This is achieved by setting a sufficiently long dead-time, $\tau_\text{d}$, to make the AP negligible and record the observed DCR, $r_\text{dc}$. From this, the intrinsic (or $\emph{dead-time corrected}$) DCR is given by,
\begin{equation}
r_\text{dc}^{*}=\frac{r_\text{dc}}{1 - r_\text{dc} \tau_\text{d}}.
\label{eq:drc_corr}
\end{equation}

Now, the temporal evolution of the AP is given by, 
\begin{equation}
p_\text{ap}(i)= \frac{C_i}{C_\text{d}-\sum\limits_{j=0}^{i-1}{C_j}} - r_\text{dc}^{*} \tau_{i},
\label{eq:pdf}
\end{equation}
where $\tau_{i}$ is the time-bin duration. 

The NFADs under test were the Princeton Lightwave NFADs (model no. E2G2) which have an active area of 25~$\mu$m and a series quench resistor of 500~k$\Omega$. Cooling is provided by a free-piston Stirling cooler (FPSC)~\cite{Korzh2014a}, which can provide operating temperatures in the range 140-300~K. Whilst the NFADs are placed on the cold-finger of the FPSC, enclosed inside a hermetically sealed chamber, all of the readout circuitry is placed at room temperature. Coaxial cables are used for the interconnection. The NFADs are DC-biased above their breakdown voltage and a capacitively-coupled read-out on the cathode provides the avalanche signal~\cite{Lunghi2012}. When an avalanche passes a user defined discrimination threshold, an output signal is generated and sent to the FPGA, whilst a hold-off circuit brings the bias voltage of the NFAD below the breakdown for a predetermined amount of time, $\tau_\text{d}$. After this dead-time the NFAD is re-triggered. NFADs have an intrinsic recharge time, which is given by the product of the device capacitance and the series resistor. For the devices used, the recharge time is expected to be $<100$~ns due to the very small parasitic capacitance~\cite{Jiang2011}. This sets the limit for the minimum  
dead-time achievable with the NFAD. Currently, this is constrained by the implemented read-out circuitry, meaning the minimum dead-time achieved in the following investigation was about 300~ns. The longest dead-time used in this work was around 200~$\mu$s. Given this range of available dead-times and the possibility of operating with a very low DCR (down to about 1~cps), we are able to investigate the AP evolution for a temporal range of over 3 orders of magnitude.

\begin{figure}
\begin{center}
\includegraphics[width=0.6\textwidth]{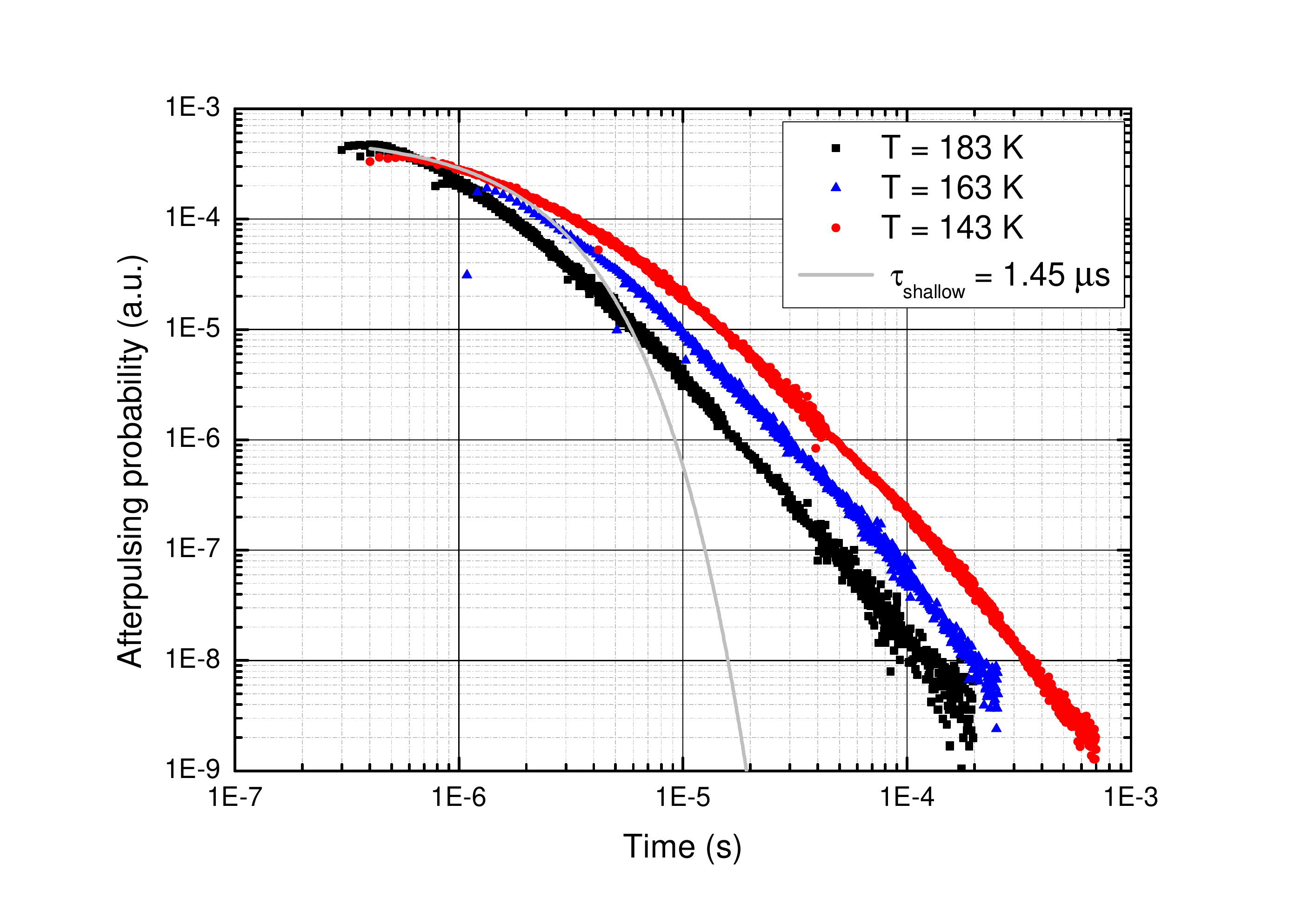}
\end{center}
\caption{Low temperature data set. The curves represent the temporal distribution of the afterpulsing probability at temperatures of 143~K, 163~K and 183~K.}\label{fig:low}
\end{figure}
\begin{figure}
\begin{center}
\includegraphics[width=0.6\textwidth]{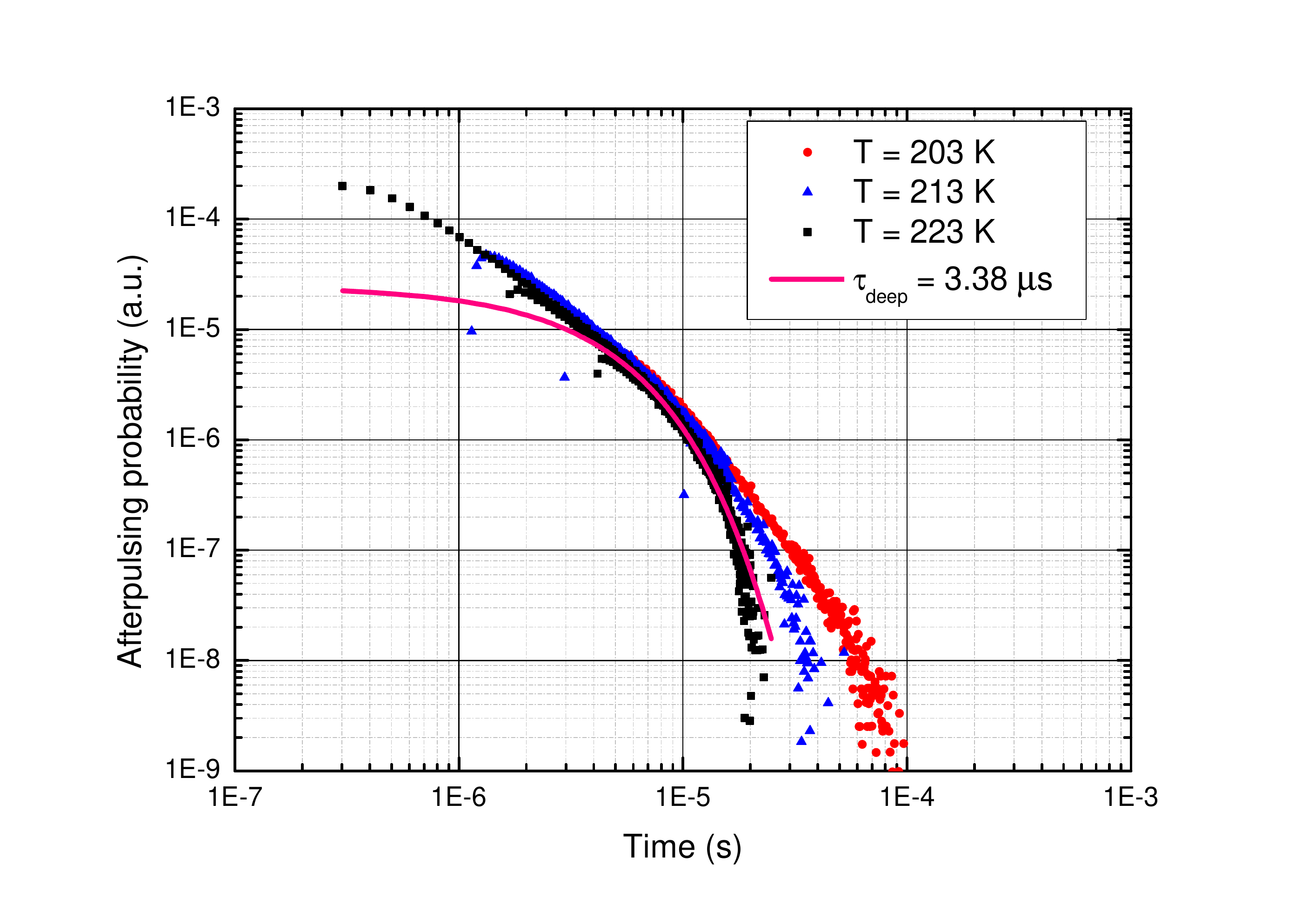}
\end{center}
\caption{High temperature data set. The curves represent the temporal distribution of the afterpulsing probability at temperatures of 203~K, 213~K and 223~K.}\label{fig:high}
\end{figure}

\section{Alfterpulse temporal behavior at different temperatures}\label{sec:ap_results1}

In this section we present the experimental characterization of the AP of a commercial NFAD over a wide temperature and temporal range. This data can be used to support the hypothesis discussed in Sec.~\ref{sec:model1}, suggesting that the the discrete trap-level model is ineffective in describing the AP evolution in time.

The collected time-resolved afterpulsing curves are divided into two distinct groups that we can identify as a high temperature and a low temperature data set. Figure~\ref{fig:low} shows the low temperature data set where the AP is measured for three different temperatures, 143~K, 163~K and 183~K. The results do not agree well with the discrete model since even at the lowest temperature of 143 K the curve is best fitted with a $t^{-2}$ trend, meaning there must be more than one trap family active, however there is no evidence of a step-like trend. As explained in Sec.~\ref{sec:model1}, this implies that the trap families involved must have an energy difference of less than 0.03~eV between them. \del{Moreover, the slope of the curve remains constant also at increasing temperatures. This clearly suggest that the afterpulsing effect can be best described by a continuous spectrum of trapping levels rather than a discrete set of trapping centers since their lifetime would change with temperature, resulting in a different slope of the curve.}

Another interesting feature of Fig.~\ref{fig:low} is the change of slope present for each curve at a different time. The curve prior to this transition time, which we name $t_{\text{knee}_1}$, can be explained as the exponential contribution of the trap family with the shallowest energy level. In fact, as introduced in Sec.~\ref{sec:model1}, for sufficiently low temperatures, if the distribution of trap levels has an upper boundary, it can be isolated as an exponential contribution to the AP distribution at short times. Since the lifetime of the shallowest trap has an exponential dependence with temperature, also $t_{\text{knee}_1}$ will show this dependence and will eventually be masked by the dead-time of the detector at higher temperatures.

\begin{figure}
\begin{center}
\includegraphics[width=0.6\textwidth]{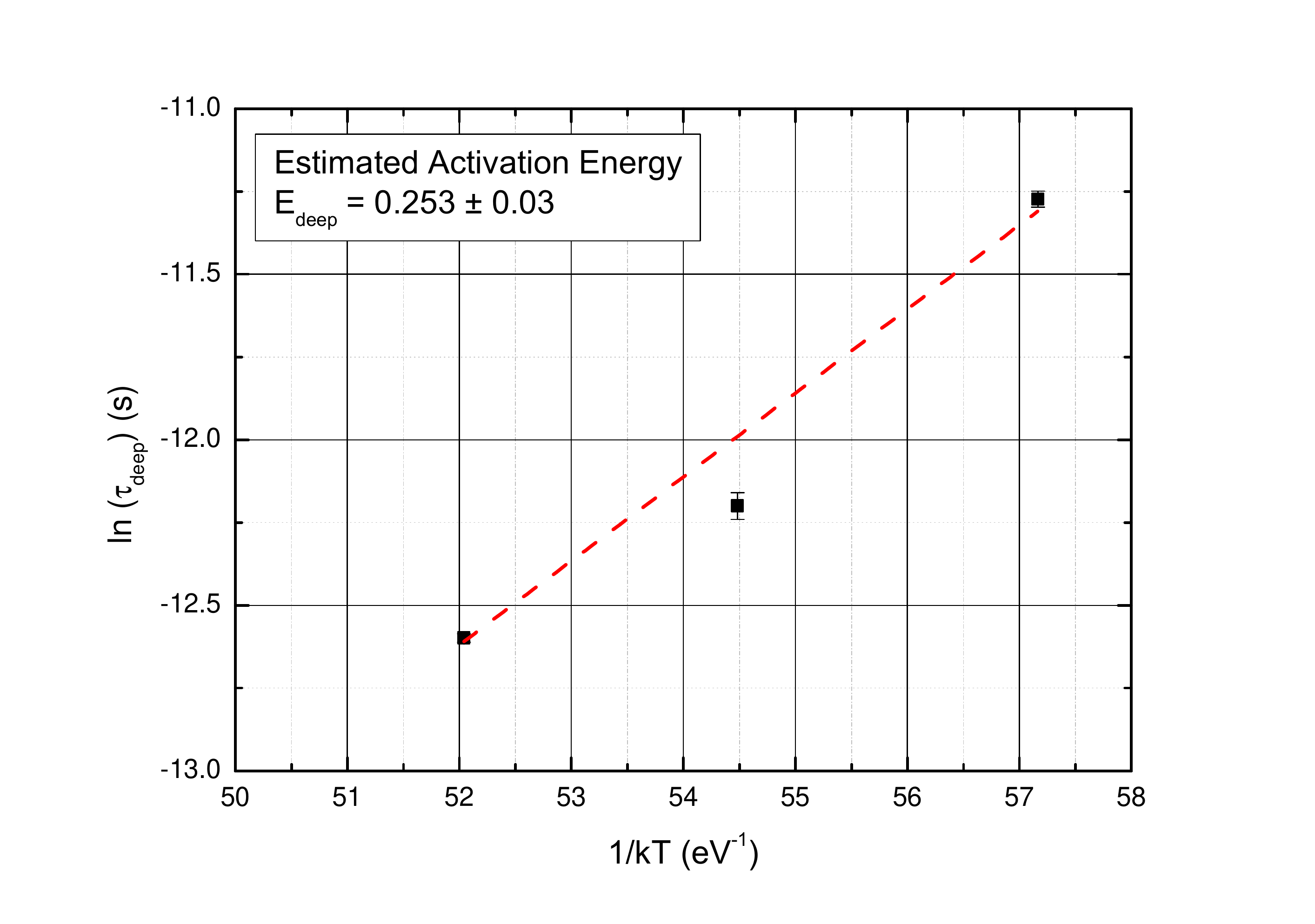}
\end{center}
\caption{Arrhenius plot extracted from the exponential fit of the last portion (after $t_{\text{knee}_2}$) of the three curves in the high temperature data set. The calculated activation energy of the deep trap is approximately 0.25~eV.}
\label{fig:arr_high}
\end{figure}

This hypothesis is also confirmed by Fig.~\ref{fig:high} which presents the high temperature data set collected at 203~K, 213~K and 223~K. Here the initial change of slope of the curve is not visible since $t_{\text{knee}_1}$ is masked by the dead-time. However, a new slope transition ($t_{\text{knee}_2}$) is present, which represents the instant after which the main contribution to the afterpulsing is the deepest trap family. We are able to fit Eq.~\ref{eq2} to this section of the high temperature data set as shown in Fig.~\ref{fig:high}. The extracted de-trapping constants, $\tau_\text{deep}$, can be represented on an Arrhenius plot (see Fig.~\ref{fig:arr_high}) to extract the activation energy of the deepest trap family. We find $E_\text{deep} = 0.253 \pm0.00 3$~eV.

These results open new questions on the physical modeling of the afterpulsing effect: the discrete trap model is clearly not adequate for explaining the experimentally observed AP distributions, however, even more complex models presented in literature have so far not captured all of the observation, whilst maintaining a sound physical interpretation. In Ref.~\cite{Itzler2012} the authors proposed a model based on a continuous distribution of de-trapping rates, which gave rise to a $t^{-1.2}$ dependency and fitted a large set of afterpulsing curves. Unfortunately, all curves presented were at temperatures higher than 200~K and were limited to a temporal range smaller than 1~$\mu$s. Recently, another study~\cite{Horoshko2014} presented results similar to our findings but for Silicon SPADs and within a limited temporal range and at only one temperature. In the following section we discuss a possible extension of the AP model which encompasses the findings of this section and aims to shed light on the $t^{-2}$ trend observed between the times $t_{\text{knee}_1}$ and $t_{\text{knee}_2}$. These consideration will allow us to estimate the activation energy of the shallowest trap family, $E_\text{shallow}$.
\begin{figure}
\begin{center}
\includegraphics[width=0.6\textwidth]{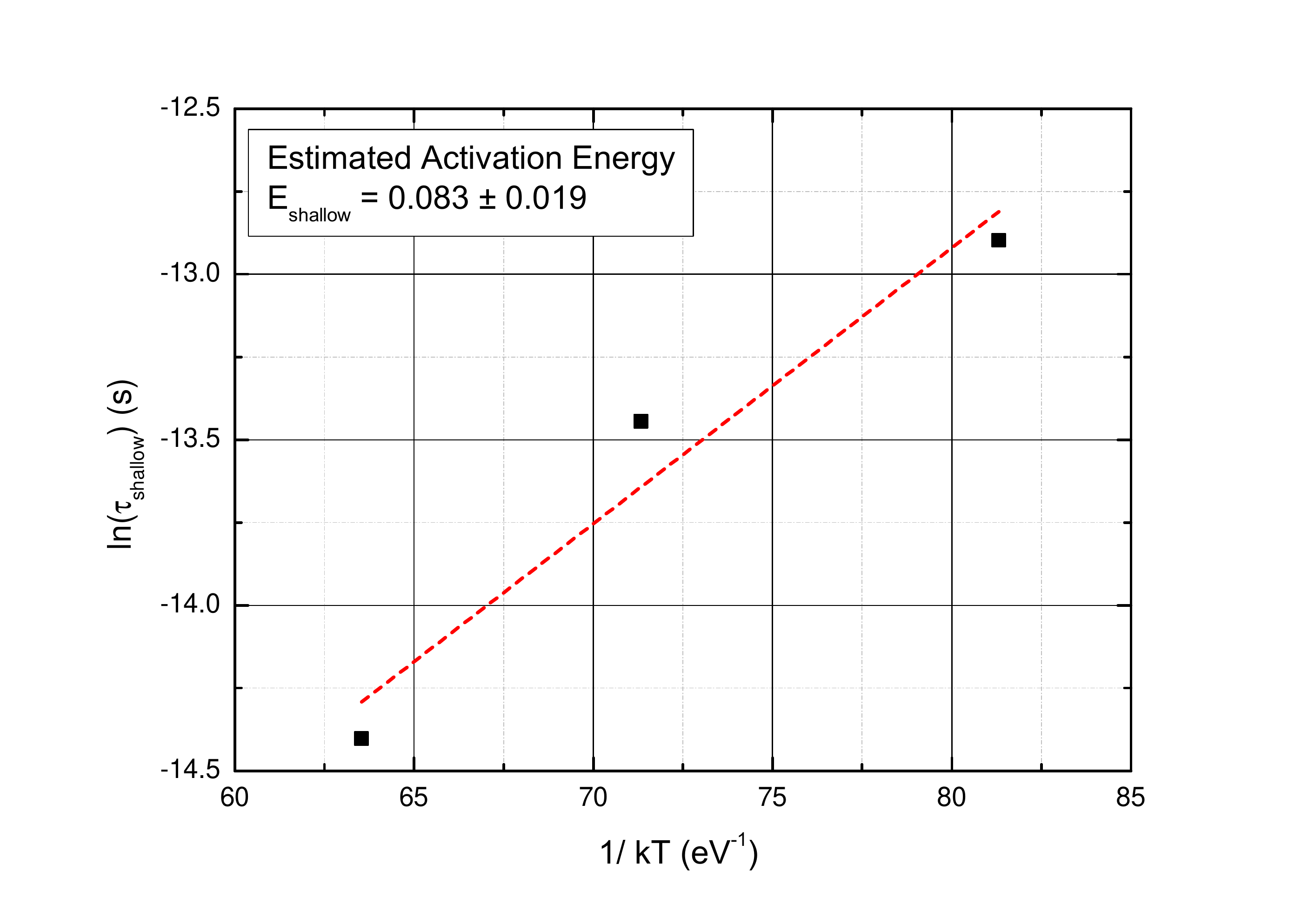}
\end{center}
\caption{Arrhenius plot extracted from the exponential fit of the first portion (before $t_{\text{knee}_1}$) of the three curves in the low temperature data set. The calculated activation energy of the shallow trap is approximately 0.08~eV.}\label{fig:arr_low}
\end{figure}


\section{Considerations for an extended afterpulsing model}\label{sec:ap_model2}

In order to consider a continuous spectrum of traps we need to rewrite Eq.~\ref{eq1} in an integral form, obtaining
\begin{equation}\label{eq5}
p_{ap}(t) = \int_{E_{shallow}}^{E_{deep}} A \frac{1}{\tau(E, T)} e^{-\frac{t}{\tau(E, T)}} dE.
\end{equation}
It is easier to evaluate this formula using the de-trapping rate $R = 1/\tau(E, T)$~\cite{Itzler2012}. Since $E \propto \text{ln}(R)$ we need to make the substitution $dE \propto \frac{1}{R}dR$. This gives
\begin{equation}\label{eq3}
p_{ap}(t) = \int_{R_{deep}}^{R_{shallow}} A e^{-tR} dR,  
\end{equation}
where $R_{deep}$ and $R_{shallow}$ are the de-trapping rates of the deepest and the shallowest trap families. The term $A$ is the capture probability and corresponds to 
\begin{equation}
A = n \sigma v_{th}, 
\end{equation}
where $n$ is the number of photo-generated holes, $\sigma$ is the cross section and $v_{th}$ is the speed of the holes in the material. 
Usually, $A$ is considered to be independent of the trap activation energy and temperature, however, this would ultimately lead to a $t^{-1}$ trend of the afterpulsing decay, which disagrees with our data.
Therefore we consider that $A$ could depend on the activation energy of the trap. Intuitively, $n$ and $v_{th}$ should not depend on the activation energy since these quantities are not related to the defect sites themselves. On the contrary, the cross section might depend on the energy as well as the temperature~\cite{Tapster83}. Based on this, we can obtain the $t^{-2}$ trend by assuming that the probability of capturing a carrier present in the conduction band is identical to the probability for the carrier to be released, i.e. 
\begin{equation}\label{eq4}
\sigma (E, T)=\tau_0 R.
\end{equation}
By inserting Eq.~\ref{eq4} into Eq.~\ref{eq3} and integrating, we find a fitting formula capable of modeling the afterpulse decay for all temperatures. At the lowest temperatures, the de-trapping rates for the deepest trap family are consistent with the results obtained in Sec.~\ref{sec:exp_setup}. Moreover, we now have information about the shallowest trap family which enables us to fit the low temperature data set to find the trap lifetimes. Figure~\ref{fig:low} shows the fitted curve for the 143~K results. We produce an Arrhenius plot of the lifetimes in Fig.~\ref{fig:arr_low}. From this we can deduce an activation energy for the trap family closest to the conduction band, $E_\text{shallow} = 0.083\pm0.019$~eV.

\section{Conclusion}

In this article we have studied the temporal evolution of the AP over a range of times between $\sim$300~ns to $\sim$1~ms. The results show a power-law dependence over a large temporal range. However, at low temperatures, we find that the AP dependence is exponential at short delays before following a power law. On the contrary, at higher temperatures, we observe exponential behavior at long delays. The power-law dependence is consistent with the idea that there exists a dense spectrum of trap levels. In this picture, the exponential behavior at short and long delays corresponds to the de-trapping rate of defects at the edges of the energy spectrum. We define the energy difference between the conduction band and the outermost trap levels as $E_\text{shallow}$ and $E_\text{deep}$, which we experimentally find to be 0.08~eV and 0.25~eV, respectively. Future work is needed to better understand the carrier capture probability dependence on the activation energy of the trap and its temperature dependence, which is required in order to explain the precise slope of the power-law dependency. 

\section*{Acknowledgements}

We would like to thank Mark Itzler and Anthony Martin for useful discussions. This work was supported by the Swiss NCCR QSIT project.

\end{document}